\documentstyle[preprint,aps]{revtex}

\font\big=cmb10 scaled\magstep3

\begin{document}
\title{Fluctuations in Hadronic and Nuclear Collisions\thanks{%
Invited paper to the special issue of {\it Foundation of Physics }dedicated to
Mikio Namiki's 70th. birthday.}}
\author{Yogiro Hama$^{(1)}$, Takeshi Kodama$^{(2)}$\ and\ Samya Paiva$^{(1)}\thanks{%
Funda\c {c}\~{a}o de Amparo \`{a} Pesquisa do Estado de S\~{a}o Paulo
(FAPESP) fellow.}$}
\address{$^{(1)\text{ }}$Instituto de F\'{\i}sica, Universidade de S\~{a}o Paulo
C.P.66318, 05389-970 S\~{a}o Paulo-SP, Brazil \\
$^{(2)}$ Instituto de F\'{\i}sica, Universidade do Rio de Janeiro C.P.68528,
21945-970 Rio de Janeiro-RJ, Brazil}
\maketitle

\begin{abstract}
We investigate several fluctuation effects in high-energy hadronic and
nuclear collisions through the analysis of different observables. To
introduce fluctuations in the initial stage of collisions, we use the
Interacting Gluon Model{\rm \ }(IGM) modified by the inclusion of the impact
parameter. The inelasticity and leading-particle distributions follow
directly from this model. The fluctuation effects on rapidity distributions
are then studied by using Landau's Hydrodynamic Model in one dimension. To
investigate further the effects of the multiplicity fluctuation, we use the
Longitudinal Phase-Space Model, with the multiplicity distribution
calculated within the hydrodynamic model, and the initial conditions given
by the IGM. Forward-backward correlation is obtained in this way.
\end{abstract}

\section{Introduction}

One of the main characteristics of the high-energy hadronic or nuclear
collisions is the existence of large event-by-event fluctuations:
fluctuation of multiplicity, of particle species, of inelasticity, of
momentum distribution, of impact parameter, and so on. Usually, one tries to
describe an average trend of such a phenomenon, of multiparticle production,
by using for example hydrodynamic models \cite{Landau}\footnote{%
The field-theoretical foundation of hydrodynamic model has first been given
by M. Namiki and C. Iso\cite{Namiki1}} and with success. However, if one
analyzes the data more closely, one observes that in a given experimental
setup, even under the same initial conditions of colliding objects, events
with different final state configurations take place. This fluctuation has
either a quantum mechanical or a statistical origin or even simply
associated with the impact parameter. The so-called inclusive data for the
final-particle distributions are the averages over such event-by-event
fluctuations for a given set of experimental initial conditions. In the
usual application of hydrodynamic models, to describing the inclusive data
we presumably expect, by means of a sort of ergodic assumption {\cite
{Namiki2}, }that the average over event-by-event fluctuations is replaced by
a statistical ensemble average. However, not all the average of physical
fluctuations can be expressed in terms of the above average over statistical
ensemble of the constituent configurations. For example, the
impact-parameter and quantum-mechanical fluctuations that occur in the
initial condition of each event can never be averaged out with the use of
the ergodic hypothesis. Besides, a description in terms of the average
quantities is clearly not satisfactory in treating such quantities as
multiplicity, inelasticity, semi-inclusive rapidity distributions,
correlations among particles, etc. The main aim of this work is to discuss
the effects of such fluctuations on the observed quantities and to present
our attempt to include them in the description of the data.

In the following, we shall give in the next two sections a brief account of
a modified version \cite{PHK,HP} of the Interacting Gluon Model (IGM) \cite
{IGM}, which will be used to generate the event-dependent fireballs. In
Section II, we discuss how the impact-parameter fluctuation can be treated
and in Section III, how IGM incorporates the energy and momentum
fluctuations of the partons to obtain the event-dependent energy and
momentum of the central fireball. Comparisons with some data, showing the
effects of these are also performed there. These fluctuations are
nevertheless not enough, for there are many data which require consideration
of other kinds of fluctuations. In Section IV, we describe how the
multiplicity fluctuation may be treated, concentrating especially in $pp$
collisions and their multiplicity distributions. To treating the energy and
momentum fluctuation of the observed particles, once the mass of the
fireball and its multiplicity are defined, we use the longitudinal
phase-space model. This is discussed in Section V. It is shown that these
two kinds of fluctuations are essential in describing the forward-backward
correlations. Conclusions are summarized in Section VI.

\section{Impact-parameter fluctuation}

In any collision process between particles or nuclei, the impact parameter
cannot be fixed {\it a priori}. This is in part due to the quantum
mechanical uncertainty, but, even if we could theoretically define the
trajectories of the incident particles like in heavy-ion collisions where
the incident objects are nearly classical, it would equally not be possible
in practice due to the actual experimental conditions. We may recall that
there exist some experimental techniques to discriminate, {\it a posteriori}%
, the central from the peripheral collisions in such reactions. But they do
not eliminate ponderable uncertainties. So in any realistic description of
hadronic or nuclear collisions the impact-parameter fluctuation must be
included. In previous works \cite{PHK,HP}, we have studied this fluctuation
in connection with IGM and hydrodynamic model and shown that it affects the
observables such as the inelasticity, leading-particle spectra, rapidity
distributions of produced particles in a significant amount.

The impact parameter $\vec{b}$ defines, in the first place, the {\it %
probability density of occurrence of a reaction} (apart from the
normalization) $F(\vec{b})=1-|S(\vec{b})|^2$, where the eikonal function is
written as 
\begin{equation}
|S(\vec{b})|^2=\exp \{-C\!\int d\vec{b}^{\prime }\,\!\int d\vec{b}^{\prime
\prime }\!\,\!D_A(\vec{b}^{\prime })\,D_B(\vec{b}^{\prime \prime })\,f(\vec{b%
}+\vec{b}^{\prime }\!-\vec{b}^{\prime \prime })\}=\exp \{-C\,h_{AB}(\vec{b}%
)\},  \label{S(b)}
\end{equation}
where $D_{A,B}(\vec{b})$ are the thickness functions of the incident
particle (nucleus) $A$ and $B$ respectively. $C$ is an energy-dependent
parameter and, though not necessarily so, in studying $pp$, $p$-nucleus and
nucleus-nucleus collisions, we assume it to be universal, so that it may be
determined by the condition $\int F_{pp}(\vec{b})\,d\vec{b}=\sigma
_{pp}^{inel}(\sqrt{s})$ for $pp$ collision. Notice that, because of this,
once $pp$ cross-section is fixed, the $AB$ cross-section $\sigma
_{AB}^{inel}(\sqrt{s})=\int F_{AB}(\vec{b})\,d\vec{b}$ may be calculated by
using (\ref{S(b)}). We take, as an input {\cite{ppXsection},} {\ } 
\begin{equation}
\sigma _{pp}^{inel}=56\,(\sqrt{s})^{-1.12}+18.16\,(\sqrt{s})^{0.16}.
\end{equation}
The function $f(\vec{b})$ in (\ref{S(b)}), subject to the constraint $\int f(%
\vec{b})\,d\vec{b}=1,$ accounts for the finite {\it effective} parton
interaction range (with the screening effect taken into account). The
simplest choice of $f(\vec{b})$ would be the point interaction $\delta (\vec{%
b})$, but we prefer to parametrize it as a Gaussian with a range $\approx
0.8\;fm$, which is more consistent with the character of the strong
interaction and also describes better the data. For proton, we parametrize $%
D_p(\vec{b})$ as a Gaussian distribution. So, we have eventually $D_p(\vec{b}%
)=f(\vec{b})=(a/\pi )\;{\rm exp}(-ab^2)$, with $a=3/(2\,R_p^2)\,$, where $%
R_P\approx 0.8\;fm$ is the proton radius. For nuclei, we take 
\begin{equation}
D_A(\vec{b})=\int_{-\infty }^{+\infty }\!\!\!\rho _A(\vec{b}%
,z)\,dz={}\int_{-\infty }^{+\infty }\!\!\frac{\rho _0}{1+\exp \left[ \left(
r-R_0\right) /d\right] }dz\,,
\end{equation}
where $R_0=r_0A^{1/3}$, $r_0=1.2$ fm\thinspace , $d=0.54$ fm and $\rho _A(%
\vec{r})$ is normalized to $A$\thinspace . Thus, in the particular case of $%
pA$ collisions, we get 
\begin{equation}
h_{pA}(\vec{b})=a\int_0^\infty db^{\prime }\,b^{\prime }\,D_A(\vec{b}%
^{\prime })\,I_0(a\,b\,b^{\prime })\,e^{-{a}(b^2+b^{\prime }{}^2)/2}\,,
\label{h(b)}
\end{equation}
where $I_0$ is a modified Bessel function.

Besides, the impact parameter determines {\it the size of the fireball},
because as $b$ increases the overlap of the hadronic matter becomes smaller
and consequently so does the average mass of the fireball generated in the
collision. We incorporate this effect by writing the parton momentum
distribution functions as 
\begin{equation}
G_A(x,\vec{b})=D_A(\vec{b})\,/x\,,\;\;\;\;\;\;G_B(y,\vec{b})=D_B(\vec{b}%
)\,/y\,,  \label{G}
\end{equation}
where $x$ and $y$ are the Feynman variables of partons in $A$ and $B$,
respectively, in the equal-velocity (e.v.) frame. With this, we are assuming
that the parton momentum distribution is independent of the particular type
of nucleus and also equal for whole the nucleus, the only difference being
their density (where the thickness $D(\vec{b})$ is large, more partons of a
given momentum are found). Here, it is presumed that {\it the same physics
describes $pp$, $pA$ and }$AB${\it \ collisions} and that {\it no
correlation exists among the nucleons inside each of the incident nuclei}.
Then, given an impact parameter $\vec{b}$, the density of parton pairs with
momenta $(x,-y)\sqrt{s}/2$ that fuse contributing to the fireball formation
is written as 
\begin{eqnarray}
w(x,y;\vec{b}) &=&\int d\vec{b}^{\prime }\,\int d\vec{b}^{\prime \prime
}\,\,G_A(x,\vec{b}^{\prime \prime })\,G_B(y,\vec{b}^{\prime \prime
})\,\sigma _{gg}(x,y)\,f(\vec{b}+\vec{b}^{\prime }-\vec{b}^{\prime \prime
})\;\theta \left( xy-M_{\min }^2\,/s\right)  \nonumber \\
&=&h_{AB}(\vec{b})w(x,y)\,,  \label{W(xyb)}
\end{eqnarray}
with 
\begin{equation}
w(x,y)=\left[ \sigma _{gg}(x,y)\,/xy\right] \,\theta \left( xy-M_{\min
}^2\,/s\right) ,  \label{W(xy)}
\end{equation}
where $M_{\min }=2m_\pi $ and the parton-parton cross-section is
parametrized as\footnote{%
In \cite{PHK}, following the original IGM \cite{IGM}, we had parametrized (%
\ref{W(xy)}) with $\sigma_{pp}^{inel}$ in the denominator. In \cite{HP}, we
have redefined $\alpha$ as an adimensional constant like in this expression.}
$\sigma _{gg}(x,y)=\alpha \,/\left( xys\right) $.$\,$Observe that in (\ref
{W(xyb)}), $\vec{b}$ dependence is factorized out.

\section{Energy and momentum fluctuation of the central fireball:
Interacting-Gluon Model}

Interacting Gluon Model (IGM) \cite{IGM} is a simple QCD motivated model,
especially designed to create the initial conditions for hydrodynamic
descriptions, by incorporating in an intuitive way the microscopic
fluctuations in the initial stage of the collision. It is based on an idea {%
\cite{VanHove}} that in high-energy collisions valence quarks weakly
interact so that they almost pass through, whereas gluons interact strongly,
producing an indefinite number of mini-fireballs, which eventually form a
unique large central fireball (all possible $q\bar{q}$ sea quarks are, in
this model, ``converted'' to equivalent gluons). One of the nice features of
this model is its easy handling. However, the main drawback of the original
version was the neglect of the impact-parameter. In \cite{PHK,HP}, we have
improved it, by including the impact-parameter fluctuation as explained in
the previous section. By reinterpreting the partons in that section as
gluons, since only these are assumed to interact, the probability density $%
\chi (E,P;\vec{b})$ of forming a fireball with energy $E$ and momentum $P$
at a fixed $\vec{b}$ is obtained as follows. We assume that the colliding
objects form a fireball, via gluon exchanges, depositing in it momenta $x(%
\vec{b})\sqrt{s}/2$ and $-y(\vec{b})\sqrt{s}/2$, respectively. Let $n_i$ be
the number of gluon pairs that carry momenta $x_i\sqrt{s}/2$ and $-y_i\sqrt{s%
}/2$. Thus, 
\begin{equation}
\sum_in_ix_i=x(\vec{b})\;\;\;\;\;\;\mbox{and}\;\;\;\;\;\;\sum_in_iy_i=y(\vec{%
b})\,.  \label{x_and_y}
\end{equation}
In what follows, we will omit the explicit $\vec{b}$ dependence of $x$ and $%
y $ in order not to overload the notation. The energy and momentum of the
central fireball in the e.v. frame of the incident particles are given by 
\begin{equation}
E=(x+y)\,\sqrt{s}/2,\,\qquad P=(x-y)\sqrt{s}/2  \label{EandP}
\end{equation}
and its invariant mass $M$ and rapidity $Y$ are respectively 
\begin{equation}
M=\sqrt{sxy}\equiv \kappa \sqrt{s}\;\;\;\;\;\mbox{and}\;\;\;\;\;Y=(1/2)\ln
\left( x/y\right) \,.  \label{MY}
\end{equation}

With these notations, we can follow the prescription given in \cite{IGM} and
write the relative probability of forming a fireball with a specific energy
and momentum as 
\begin{equation}
\Gamma (x,y;\vec{b})\simeq \exp \{-{\bf X}^T{\bf G}^{-1}{\bf X}\}/\left[ \pi
\,\sqrt{\det ({\bf G})}\right] \,,
\end{equation}
where 
\[
\;{\bf X}=\left( 
\begin{array}{c}
x\!-\!\langle x\rangle \\ 
y\!-\!\langle y\rangle
\end{array}
\right) \;,\qquad {\bf G}=2\left( 
\begin{array}{cc}
\langle x^2\rangle & \langle xy\rangle \\ 
\langle xy\rangle & \langle y^2\rangle
\end{array}
\right) \;, 
\]
with the notation 
\begin{equation}
\langle x^my^n\rangle =\int dx^{\prime }\int dy^{\prime }\,x^{\prime
}{}^m\,y^{\prime }{}^n\,w(x^{\prime },y^{\prime };\vec{b})\,,  \label{<xy>}
\end{equation}
or, in terms of $E$ and $P$, 
\begin{equation}
\Gamma (E,P;\vec{b})\simeq \left[ 2\sqrt{a_1a_2}/\pi \right] \exp
\{-a_1(E-\langle E\rangle )^2-a_2P^2\},  \label{chiepfinal}
\end{equation}
where $a_1=[s(\langle x^2\rangle +\langle xy\rangle )]^{-1}$, $%
a_2=[s(\langle x^2\rangle -\langle xy\rangle )]^{-1}$ and $\langle E\rangle
=(\langle x\rangle +\langle y\rangle )\sqrt{s}/2\,$ (\underline{don't
confuse this notation with the average value}; it is not because $w(x,y;\vec{%
b})$ is not normalized). Apparently, $\Gamma (E,P;\vec{b})$ in (\ref
{chiepfinal}) is normalized. However, both $E$ and $P$ are bounded because
of the energy-momentum conservation constraint. It is also constrained by $%
M> $ $M_{{\rm \min }}$ $=2m_\pi \,$. So, we put some additional factor $\chi
_0(\vec{b})$, 
\begin{equation}
\chi (E,P;\vec{b})=\chi _0(\vec{b})\Gamma (E,P;\vec{b})\,,  \label{chi}
\end{equation}
such that 
\begin{equation}
\int \!dP\int \!dE\;\chi (E,P;\vec{b})\;\theta (\sqrt{E^2-P^2}-M_{{\rm \min }%
})=\frac{F_{AB}(\vec{b})}{\sigma _{AB}^{inel}}\,.  \label{chinormal}
\end{equation}

As implied by (\ref{G}) and remarked there, the gluon momentum distribution
is independent of the particular type of nucleus, the only difference being
their density. So, in the integral (\ref{<xy>}), $x^{\prime }$ and $%
y^{\prime }$ vary from some lower limit, defined by $\sqrt{sx^{\prime
}y^{\prime }}=M_{{\rm \min }}\,$, up to $1$, corresponding to the complete
neglect of any collective effect of the nucleons in a nucleus. On the other
hand, the integration limits of (\ref{chinormal}) are chosen differently. $x$
and $y$ in (\ref{x_and_y}) may be larger than $1$, because gluons from
different nucleons may contribute to give the fireball a momentum transfer
that is larger than $\sqrt{s}/2$\thinspace , which is just the incident
momentum of a single nucleon in our e.v. frame. We take as the upper limit
of $x$ and $y$ the overlap $h_{A,B}(\vec{b})$, whenever it is larger than $1$%
, and limited to $A$ and $B$, respectively ($x$ or $y$ remains $\leq 1\,$,
if it corresponds to a proton). When $h_{A,B}(\vec{b})<1$, we take it $=1$,
because in such a case just a single nucleon of each nucleus interacts.

\subsection{Inelasticity distributions}

The concept of inelasticity, understood as the fraction of the incident
energy $E_0$ which is lost while a particle interact with another one, is
crucial in cosmic-ray data analysis where the primary mass composition,
hence an important information about the Universe, is deduced by using
models of cascade development with appropriate inelasticity distributions%
\footnote{%
For a recent comparative study of the main existing models on inelasticity,
see for example \cite{Wilk}.} and cross sections as the inputs. In this
case, the usual definition is $k=(E_0-E^{\prime })/E_0$\thinspace , where $%
E^{\prime }$ is the leading (or surviving) particle energy. At high energy,
a nucleus hardly supports any collision without suffering a breakup, so the
term inelasticity in this sense is commonly reserved to a hadron projectile
and, in this paper, we shall restrict the consideration only to proton
incident on nuclear targets. However, in a wider sense, as the fraction of
the incident energy $E_0$ which is materialized into produced particles, it
is also important in connection with the production of a quark-gluon plasma
in heavy-ion collisions \cite{Busza,Hwa,Wong,Hufner,Csernai,IGM}. In Ref.%
\cite{IGM}, $\kappa $ appearing in (\ref{MY}) is called inelasticity and
this is the quantity of interest in heavy-ion collisions. In the present
paper, we shall adopt the usual definition $k$ given above and, as for $%
\kappa $, call it simply $\kappa $. In any case, it is clear that {\it we
can talk about inelasticity distribution only when there exists some
event-by-event fluctuation}.

Having obtained $\chi (E,P;\vec{b})$\thinspace , we can readily compute such
distributions. The $\kappa $-distribution has been obtained in \cite{PHK}
and reads 
\begin{equation}
\chi (\kappa )=\int d\vec{b}\int dE\int dP\,\chi (E,P;\vec{b})\,\delta (%
\sqrt{(E^2-P^2)/s}-\kappa )\,\theta (\sqrt{E^2-P^2}-M_{{\rm \min }})\,.
\label{chi_K}
\end{equation}
Then, by fitting the only existing $\chi (\kappa )$ data \cite{Brick} at $%
\sqrt{s}=16.5$ GeV, we fix the parameter $\alpha $ of the model as $\alpha
=21.35$. A comparison with the data is shown in Fig.1, where we have also
put the result of \cite{IGM}. It is seen that the original version of IGM
already gives a reasonable description of the data, but the inclusion of the
impact-parameter fluctuation drastically improves the agreement. The latter
enhances the small-$\kappa $ events and makes the overall shape flatter. The
enhancement of large-$\kappa $ events is simply due to the larger value of $%
\alpha $ which is necessary for an overall fitting now.

The inelasticity distribution $\chi (k)$ has been obtained in \cite{HP} and
reads 
\begin{equation}
\chi (k)=\int d\vec{b}\int dE\int dP\,\chi (E,P;\vec{b})\,\delta ((E+P)/%
\sqrt{s}-k)\,\theta (\sqrt{E^2-P^2}-M_{{\rm \min }})\,.  \label{chi_k}
\end{equation}
Often, the inelasticity is defined in the lab. frame but, except when $%
k\rightarrow 1$, the difference between $k$ defined in this~frame and the
one given in the e.v.~frame is quite negligible. So, we will not make any
distinction here and compute everything in the latter. We show, in Fig.2,
the results for several $pA$ collisions at $\sqrt{s}=550$ GeV. No
accelerator data at such a high energy exists, but it is seen that $\chi (k)$
is nearly $k$ independent for $pp$, in agreement with $ISR$ data \cite
{Basile,Capiluppi}. In a recent cosmic-ray experiment \cite{Ohsawa}, hadron-$%
Pb$ inelasticity distribution at an average energy of $\langle \sqrt{s}%
\rangle =550$ GeV has been estimated. The result\footnote{%
In \cite{HP}, we cited the preliminary result communicated by E.\thinspace
Shibuya, a member of the collaboration, which is slightly different.} is $%
\chi (k)\simeq (-0.25\pm 0.50)(1-k)^{1.85\pm 1.77}+(3.1\pm 1.7)k^{1.85\pm
1.77}$. We find that the qualitative features of our result agree with this
estimate, except in the low-$k$ region. Some of the origins of the
discrepancy may be the difference between $\pi -Pb$ and $p-Pb$ collisions
and the absence of the leading-particle fragmentation in our model. However,
the uncertainty in their estimate is quite large.

We show in Fig.3 the average inelasticity $\langle k\rangle $ as function of 
$\sqrt{s}$, for several target nuclei. In Ref.\cite{Wilk}, several models
for inelasticity have been compared with their estimates obtained with
cosmic-ray data, which show slowly decreasing behavior with $E_0$\thinspace
. According to their comparison, IGM is the model which predicts the most
quick fall of the average inelasticity with $E_0$ in clear conflict with
their estimates. Our curves in Fig.3 still decrease as $\sqrt{s}$ increases
but, compared with the results of \cite{IGM}, the energy-dependence is quite
small now and compatible with the estimates obtained in \cite{Wilk}. The
main origin of this contrast is the factor $\sigma _{pp}^{inel}$ which has
been dropped out in (\ref{W(xy)}), because it is not necessary in our
version.

\subsection{Leading-particle spectra}

A related quantity is the leading-particle spectrum, as shown in Fig.4 at $%
\sqrt{s}=14$ GeV \cite{LPS}. Since data on $p_T$ dependence are scarce, we
have assumed an approximate factorization of $x_l(=2p_l/\sqrt{s})$ and $p_T$
dependences, 
\begin{equation}
E_l({d^3\sigma }/{dp^3})\approx \;f(x_l)h(p_T)\,,
\end{equation}
where 
\begin{eqnarray}
f(p_l)=\int d\vec{b}\int dP\int dE\,\chi (E,P;\vec{b})\,\theta (\sqrt{E^2-P^2%
}-M_{{\rm \min }})\;\delta (\left[ \sqrt{s}-\left( E+P\right) \right]
/2-p_l)\,,
\end{eqnarray}
and parametrized $h(p_T)$ as 
\begin{equation}
h(p_T)=(\beta ^2/2\pi )\,e^{-\beta \,p_T}\,,  \label{h(p)}
\end{equation}
determining the average $\beta $ by using the $p_T$ dependence of the data 
\cite{LPS}. The curves obtained with these $\beta $ values (with an
interpolation for $Al$ and $Ag$) are shown in Fig.4. The result of \cite{IGM}
for $pp$ is also shown for comparison. Again, it is seen that a reasonable
agreement with the data is obtained with IGM only, but the inclusion of the
impact-parameter fluctuation improves it remarkably. We did not put their
curves for the other targets, but the behavior is similar, namely they are
more bent showing a definite deviation from the data in the largest-$x_l$
region. This is a consequence of the neglect of the peripheral events there.
Some authors \cite{Hwa,Hufner,Csernai} have obtained good fits to $pA$ data,
but in those works it is not clear which is the connection to other relevant
quantities such as momentum distributions, correlations,... of the secondary
particles. Also, $pp$ is usually treated as a separate case.

\subsection{Rapidity distributions}

Let us now study the effects of fluctuations we have just introduced on the
final particle spectra. As mentioned in the Introduction, the so-called
inclusive rapidity (or pseudo-rapidity) distributions are the averages over
such event-by-event fluctuations for a given set of experimental initial
conditions. However, in usual computations of these distributions, say, by
use of hydrodynamic models, instead of taking such averages 
\begin{equation}
\left\langle \frac{dN}{dy}\right\rangle =\int d\vec{b}\,\int dP\,\int dE\,%
\frac{dN}{dy}(E,P;\vec{b})\,\chi (E,P;\vec{b})\,\theta (\sqrt{(E^2-P^2)/s}%
\,-\kappa _{{\rm min}})\;,  \label{<dN/dy>}
\end{equation}
one considers some average initial conditions and calculates the
distributions, starting from them. Namely, 
\begin{equation}
\frac{dN}{dy}\left( \left\langle E\right\rangle ,\left\langle P\right\rangle
=0,\left\langle \vec{b}\right\rangle \right) \,.  \label{dN/dy}
\end{equation}
It is evident that only under very special conditions that these two
quantities can coincide.

In \cite{PHK}, we studied the deviation of the latter from the more
realistic former distribution $\left\langle dN/dy\right\rangle $, by
adopting the one-dimensional Landau's hydrodynamic model for an ideal gas in
order to compute the rapidity distribution $dN(E,P,\vec{b})/dy$ for a
definite initial conditions. Despite all the simplifications, this model is
known to reproduce the main features of the measured momentum (or rapidity)
distributions and has advantage of having an analytical solution over the
whole rapidity range\cite{Khalatnikov}. The only inputs of the model are the
total energy, momentum and the geometrical size of the initial fireballs.
Remark that we are talking about the central fireballs and not about the
rest of the system. In the case the incident particles are nucleons, the
latter appears most frequently as leading particles. In order to avoid
additional complexities, let us consider only this case, namely $pp$
collisions.

The invariant momentum distribution of produced particles in a hydrodynamic
model is usually given by Cooper-Frye formula {\cite{Cooper}} 
\begin{equation}
E\;\frac{dN}{d\vec{p}}=\int_{\sigma (T_d)}\,f(p^\mu u_\mu )\;p^\mu \;d\sigma
_\mu \;,  \label{CoopFrye}
\end{equation}
where $\sigma (T_d)$ is a constant-temperature freeze-out hypersurface, $%
f(p^\mu u_\mu )$ is Bose-Einstein (or Fermi-Dirac) distribution, $p^\mu $ is
the 4-momentum of the emitted particle and $u^\mu $ is the 4-velocity of the
fluid. Although it is possible to use more realistic freeze-out criteria {%
\cite{freeze1,freeze2,freeze3,GHK}}, here we limit ourselves to the simplest
choice (\ref{CoopFrye}) without sophistication. This is enough for our
present purpose of studying how the initial condition fluctuations affect
the final particle spectra.

$\sigma (T_d)$ in (\ref{CoopFrye}) is determined by solving the hydrodynamic
equation 
\begin{equation}
\partial _\mu T^{\mu \nu }=0\,,\qquad {\rm with\qquad }T^{\mu \nu
}=(\varepsilon +p)u^\mu u^\nu -pg^{\mu \nu },
\end{equation}
with an appropriate equation of state ($p=\varepsilon /3$ in our case) once
the dissociation temperature $T_d$ is given. As for the ``initial volume''
for a fireball of mass $M$, in the lack of a better justification founded on
a physical basis, we adopt in the present work 
\begin{equation}
V_0=\frac{2m_p}M\,V\,,  \label{volume}
\end{equation}
which has been suggested by a phenomenological analysis {\cite{Carruthers}}
of the $M$ dependence of average multiplicity data {\cite{multiplicity}} and
also consistent with the $M$ dependence of the momentum distribution data 
\cite{Basile2}. The initial temperature $T_0$ is then computed by putting $M$
into this volume. As remarked in \cite{PHK}, nowadays we know that neither
the hypothesis of instantaneous thermalization nor the appearance of
extremely high values of the initial temperature are physically reasonable.
However, in spite of these rather {\it non-conventional} initial conditions,
many of the qualitative and the quantitative results (average multiplicity,
particle ratios, momentum distributions, $\cdots $) are surprisingly good
when compared with data. In our point of view, perhaps the equilibrium is
attained at a later time when the system has already suffered some
expansion, but then the temperature and the rapidity distributions at the
onset of the hydrodynamic regime would be approximately those of Landau's
model whose initial conditions correspond to high temperature and energy
density if extrapolated back in time. So, for any practical purpose, we can
use Landau's solution to describe the system. We emphasize, however, that
the fluctuation effects which are the central object of the present study do
not depend sensibly on such a choice.

We show in Fig.5 a comparison of the results obtained in this way for $%
<dN/d\eta >$ and $dN/d\eta $ at $\sqrt{s}=53\;$GeV. It is seen that the
rapidity or equivalently pseudo-rapidity distributions are very sensitive to
the fluctuations in the initial conditions. The peak, in the case of $%
dN/d\eta $ computed with one fireball of mass $<M>$, corresponds to the
simple-wave solution. When the fluctuations are taken into account, such a
peak is completely smoothed away. They also cause a widening and a lowering
of the distributions. Although the main purpose of this work is just to show
the influences of the fluctuations, we may also compare them with some data {%
\cite{detaexp}}. We see that the behavior of the first one is more similar
to the data than the other one and the presence of the simple-wave peaks in
each event does not invalidate the overall agreement with data.

\section{Multiplicity Fluctuation}

We have shown in the last two sections how the impact-parameter and the
energy-momentum fluctuations in the initial stage of the collision affect
some of the observables. However, there are many more quantities whose
description cannot be given only in terms of the fluctuations considered up
to this point. Even after the mass of the fireball has been defined,
quantities such as the multiplicity, particle species, their momentum
distributions, ... vary from event to event and hydrodynamic model we have
used in the last section only describes the average behavior. Under certain
conditions (constant dissociation temperature $T_d$), it does give the
moments of the multiplicity distribution \cite{moments}, so in principle
also the multiplicity distribution itself \cite{diffraction}, but not the
fluctuating events.\ 

Let us discuss in this section how the multiplicity fluctuation may be
implemented. One way of doing this is to conveniently parametrizing the
multiplicity distribution for a fixed mass $M$ and determining the
parameters by imposing certain constraints. We choose a very simple
parametrization for the multiplicity distribution 
\begin{equation}
\psi (M,z)=Az^\nu e^{-\alpha z},  \label{psi}
\end{equation}
where, as usual, 
\begin{eqnarray}
\psi (M,z) &=&\left\langle N(M)\right\rangle \,P_N(M)\,, \\
z &=&N/\left\langle N(M)\right\rangle  \nonumber
\end{eqnarray}
with $P_N(M)$ indicating the probability of a fireball of mass $M$ decaying
into $N$ charged particles, and impose the conditions 
\begin{eqnarray}
\sum_NP_N(M) &=&1\qquad \quad \,\rightarrow \frac 12\int_0^\infty \psi
(z)\,dz=1,  \label{constraints} \\
\sum_NNP_N(M) &=&\left\langle N(M)\right\rangle \;\,\rightarrow \frac 12%
\int_0^\infty z\,\psi (z)\,dz=\langle z\rangle =1,  \nonumber \\
\sum_NN^2P_N(M) &=&\left\langle N(M)^2\right\rangle \rightarrow \frac 12%
\int_0^\infty z^2\,\psi (z)\,dz=\langle z^2\rangle =\frac{\langle
N(M)^2\rangle }{\langle N(M)\rangle ^2}.  \nonumber
\end{eqnarray}
By substituting (\ref{psi}) into these equations, we obtain 
\begin{eqnarray}
\alpha &=&\frac 1{\langle z^2\rangle -1}, \\
\nu &=&\alpha -1, \\
A &=&\frac{2\alpha ^\alpha }{\Gamma (\alpha )}.
\end{eqnarray}
The moments $\left\langle z^n\right\rangle =\langle N(M)^n\rangle /\langle
N(M)\rangle ^n$ have been calculated in \cite{diffraction} and, in
particular, it is found that $\left\langle z^2\right\rangle
=1+a_2/\left\langle N(M)\right\rangle $, with $a_2=1.105$. What we can do
is, once $M$ is fixed, to produce events following this distribution by the
use of the Monte Carlo method. In doing so, we have indeed to consider also
the charge fluctuation.

The overall multiplicity distributions calculated in this way is shown in
Fig.6. As seen, the results reproduce quite well the qualitative features of
the data in all the ISR energy region. They are slightly narrower than the
data and, as the energy increases, the discrepancy becomes more pronounced,
but probably the data begins to suffer the influence of the mini-jets there.

\section{Energy and momentum fluctuation of the observed particles:
Phase-space model}

The multiplicity fluctuation, discussed in the preceding section, does not
manifest itself only in the multiplicity distribution. With the inclusion of
this fluctuation, we are considering that the momentum distribution of the
secondary particles for a given $M$ is a superposition of distributions with
different multiplicity $N$. Moreover, even with a fixed $N$, the momentum
distribution will vary from event to event. As will be shown below, there
are observables such as the forward-backward multiplicity correlation, which
depends on this kind of fluctuation. But, then we faces the following
problem: ``How to compute the rapidity distribution of a system having a
definite mass $M$ and a definite multiplicity $N$ and in an event-dependent
way?'' This question did not arise when computing the inclusive
distribution, because the hydrodynamic model does take such a fluctuation
into account, as mentioned in the preceding section. What it does not do is
to generate each fluctuating event.

We propose to use the one-dimensional phase-space model to generate these
events. First, one-dimensional because we know from the data that the
momentum distributions in high-energy is essentially longitudinal. In a
previous work \cite{partition}, we have shown that the rapidity
distributions predicted by the one-dimensional phase-space model, given $M$
and $N$, are approximately Gaussian, as in hydrodynamic model. It was also
shown that these are insensitive to a certain class of dynamical factors
introduced in the model. Although it is indeed not guaranteed {\it a priori}
that the superposition of these distributions shall give the one obtained by
the hydrodynamic model, the result of this model has the qualitative
features of that one and an advantage of being a sum of event-dependent
distributions with fixed $M$ and $N$.

Given a mass $M$ and a multiplicity $N$, the one-dimensional phase-space
model tells us that the probability of finding an event with the particles
in the longitudinal momentum intervals 
\begin{equation}
\lbrack p_i,p_i+dp_i],\;\;\;\;\;\;\;i=1,\cdots N,
\end{equation}
is given by 
\begin{equation}
d^N{\bf P}=\frac 1{R_N(M)}\,\frac{dp_1}{2E_1}\,\frac{dp_2}{2E_2}\,\cdots \,%
\frac{dp_N}{2E_N}\,\delta (\sum_{j=1}^N\,p_j)\,\delta
(\sum_{j=1}^N\,E_j-M)\,,  \label{prob}
\end{equation}
where 
\begin{equation}
R_N(M)=\int \,\frac{dp_1}{2E_1}\,\int \,\frac{dp_2}{2E_2}\,\int \cdots
\,\int \frac{dp_N}{2E_N}\,\delta (\sum_{j=1}^N\,p_j)\,\delta
(\sum_{j=1}^N\,E_j-M)\,.  \label{phasespace}
\end{equation}
The invariant one-particle distribution is then given by 
\begin{equation}
E\frac{dN}{dp}=\frac{dN}{dy}=\frac{N\,R_{N-1}(M^{\prime })}{R_N(M)},
\label{1part_dist}
\end{equation}
normalized to $N$ and where $M^{\prime }=\sqrt{(M-E)^2-p^2}$ represents the
invariant mass of the system after subtracting the observed particle. The
correct expression of the probability would be (\ref{prob}) with some
dynamical factor. In \cite{partition}, in order to simulate the hydrodynamic
motion, we have included a factor $f(y)=\alpha e^{-\beta m_T{\rm cosh}y}$
for each particle, where $y$ is the rapidity, and shown that this is
entirely irrelevant, the final result being the same.

In Fig.7, we show one-particle pseudo-rapidity distributions in $pp$
collision at 53 GeV given by (\ref{prob}). Since we have completely
neglected any dynamical factor, we could not expect to obtain a perfect
agreement with the data. However, it is seen that the qualitative features
of the data are reproduced.

\subsection{Forward-backward correlation}

One of the data, which cannot be understood without the multiplicity
fluctuation discussed in Section IV and nicely reproduced with the
longitudinal phase-space model proposed here, refers to the so called
forward-backward charged multiplicity correlation. First of all, the data
show that the charged multiplicities in the two hemispheres are very little
correlated \cite{Basile}, presenting a large fluctuation. The correlation is
usually presented as a graph of the average charged multiplicity in the
backward hemisphere $<N_b>$ as a function of the effectively observed
charged multiplicity $N_f$ in the forward hemisphere as in Fig.8.

To begin with, remark that in the usual application of the hydrodynamic
model, without any fluctuation taken into account, the graph would reduce to
a single point. The consideration of the impact-parameter and the
initial-state fluctuations, implemented through IGM as we are proposing,
improves considerably the agreement with the data. Notice that the momentum
fluctuation of the fireball is essential, otherwise the correlation would be
too large. Now, without the inclusion of the multiplicity fluctuation and
especially the momentum fluctuation of the final particles, the correlation
begins to deviate from the experimental trends for large values of $N_f$%
\thinspace , because, in that case, large $N_f$ means large $M$ with the
fireball sitting more or less in the center of mass, with a symmetrical
distribution of particles.

\section{Conclusions and Further Outlooks}

We have investigated, in this paper, effects of several kinds of
fluctuations which appear in hadronic and nuclear collision, analyzing
different kinds of observables. The Interacting Gluon Model, improved by the
inclusion of the impact-parameter fluctuation and complemented by an
appropriate hydrodynamic model seems to describe well the the bulk of the
phenomenon, such as the cross-section, inelasticity, leading-particle
spectrum, average multiplicity and the inclusive momentum distribution.
Other quantities depend explicitly on the multiplicity and the
final-particle momentum fluctuations. In this paper, we have treated the
first one by considering the thermodynamics of the fireball and then
parametrizing the multiplicity distribution for each mass $M$ in a
convenient way. More microscopic description of this fluctuation would also
be possible. We have treated the momentum fluctuation of the final particles
by using the one-dimensional phase-space model. Although not completely
satisfactory, this tretement has shown the importance of such a fluctuation
in an event-dependent basis.

There are many other properties which clearly depend on fluctuations. One of
these quantities is the so-called semi-inclusive rapidity distribution,
namely, distribution with a fixed multiplicity interval. In principle, it
would be possible to obtain this distribution with the ingredients we have
considered here, but there is something which is missing. Especially in the
low multiplicity intervals, the effects of the diffractive processes cannot
be neglected. We are studying how to incorporate the diffractive processes
in IGM, in a consistent way. Also, to compare with the data, a treatment
which is somewhat more realistic than (\ref{prob}) is required. Another
interesting quantity that could be studied, which certainly depends on
event-by-event fluctuations we have discussed, is the Bose-Einstein
correlations of produced particles, or the so-called Hanbury-Brown Twiss
effect \cite{HBT}, frequently used in heavy-ion collisions to infer about
the space-time developpment of the hadronic matter which is formed in such
collisions. Suggestions for such a study has been given by M. Namiki {\it et
al.} in \cite{Namiki2}.

In this paper, we have completely neglected the fragmentation of the leading
particles, which certainly give non-negligible contributions to the
semi-inclusive rapidity distributions mentioned above, especially when the
multiplicity is small. In the case of nucleus-nucleus collisions, how to
treat the fragmentation of the leading nuclei is a completely open question.

\bigskip

\bigskip

\noindent{\bf Acknowledgments}

This work has been supported in part by Funda\c {c}\~{a}o de Amparo \`{a}
Pesquisa do Estado de S\~{a}o Paulo (FAPESP) under the contract 95/4635-0
and by MCT/FINEP/CNPq (PRONEX) under the contract 41.96.0886.00. We thank
E.\thinspace Shibuya for bringing the new cosmic-ray data to our knowledge.

\newpage

\noindent{\big Figure Captions}

\begin{description}
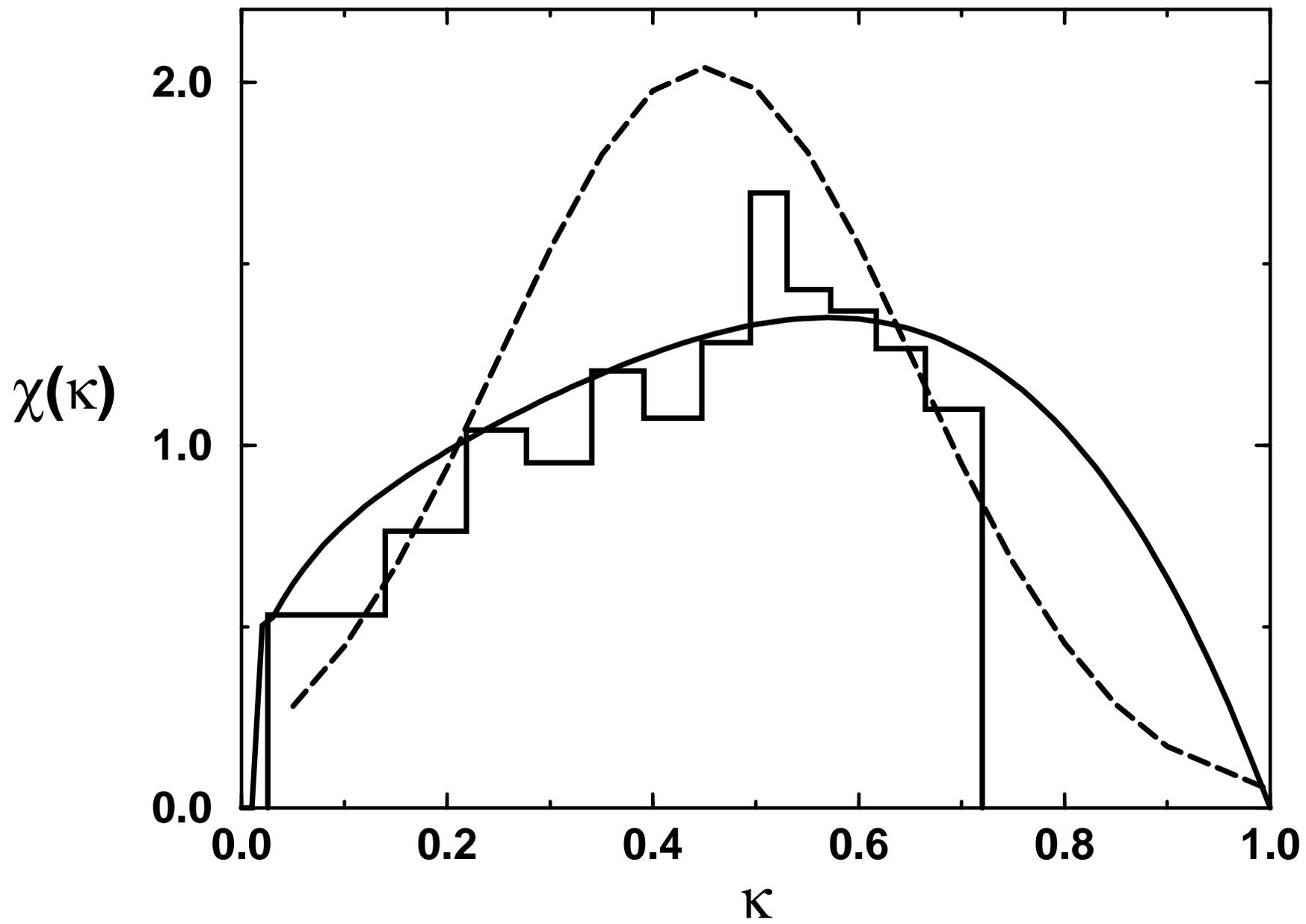
\item[Fig.1: ]  $\kappa $-distribution for $pp$ at $\sqrt{s}=16.5$ GeV. The
data are from \cite{Brick}. The solid line is our result, whereas the dashed
one is from \cite{IGM}.

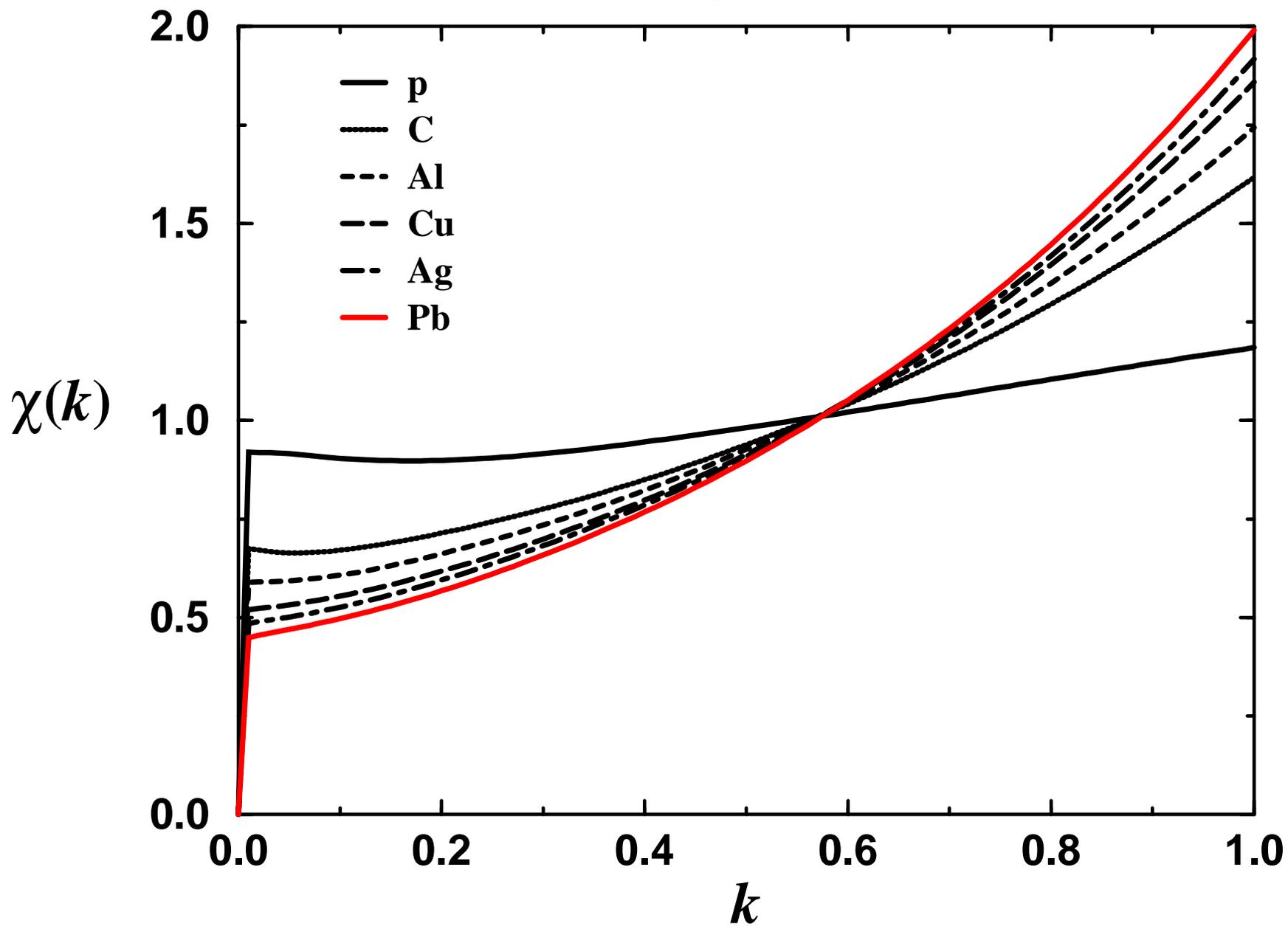
\item[Fig.2: ]  Inelasticity distribution for $pA$ collisions with several
targets at $\sqrt{s}=550$ GeV.

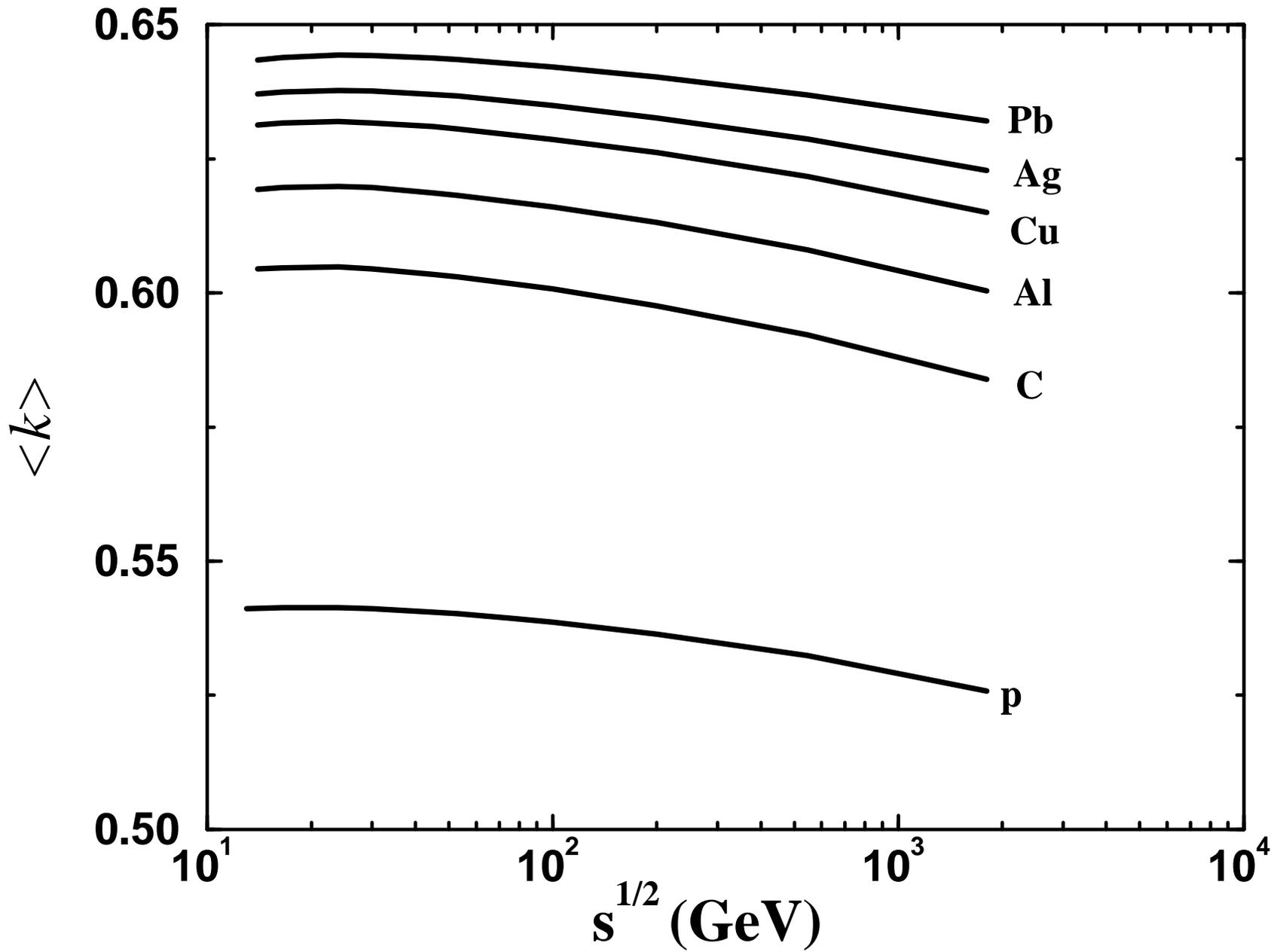
\item[Fig.3: ]  Energy dependence of the average inelasticity for $pA$
collisions.

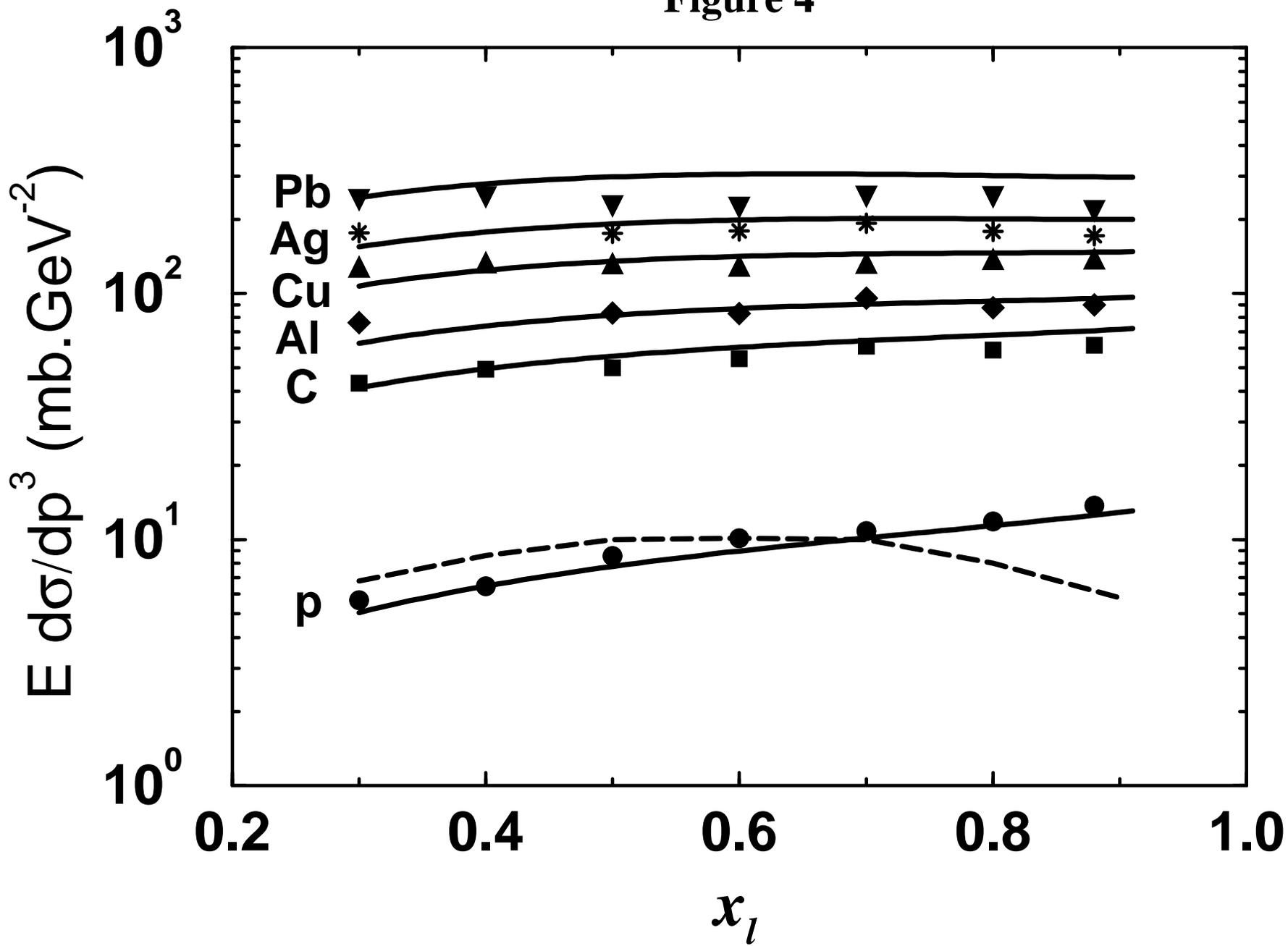
\item[Fig.4: ]  Leading-particle spectra as fuction of $x_l$ at $p_T=.3$
GeV. The data are from \cite{LPS} at $\sqrt{s}=14$ GeV. The solid curves are
our results, whereas the dashed one is from \cite{IGM}. The slope parameter
has been extracted from \cite{LPS} as $\beta =4.20,3.22,3.21,3.26,3.47$ and $%
3.78{\rm GeV}^{-1}$ for $p$, $C$, $Al$, $Cu$, $Ag$ and $Pb$ targets,
respectively.

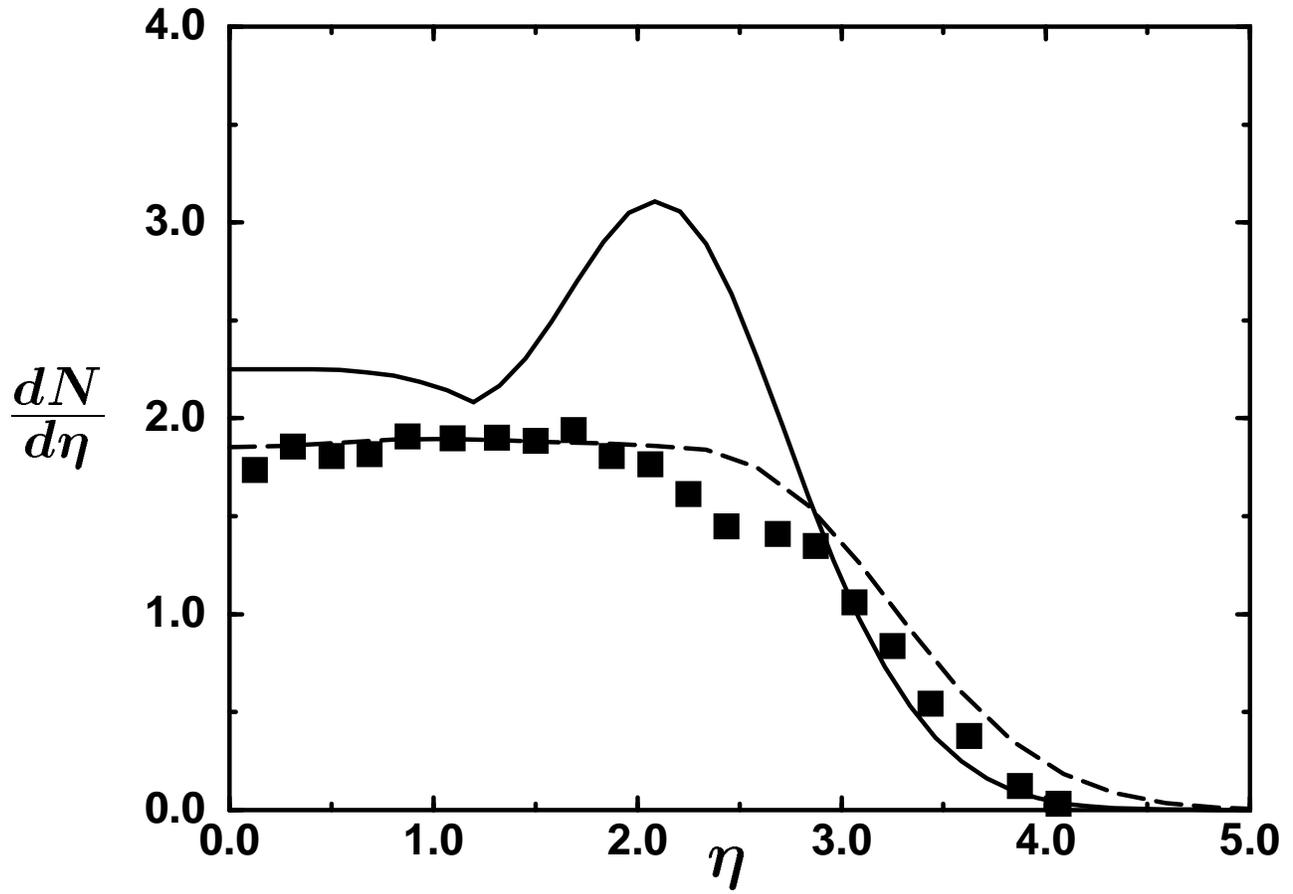
\item[Fig.5: ]  Pseudo-rapidity distributions calculated in the usual
procedure (solid line) and with fluctuations (dashed line) at $\sqrt{s}=53\;$%
GeV. Experimental data \cite{detaexp} are shown for comparison.

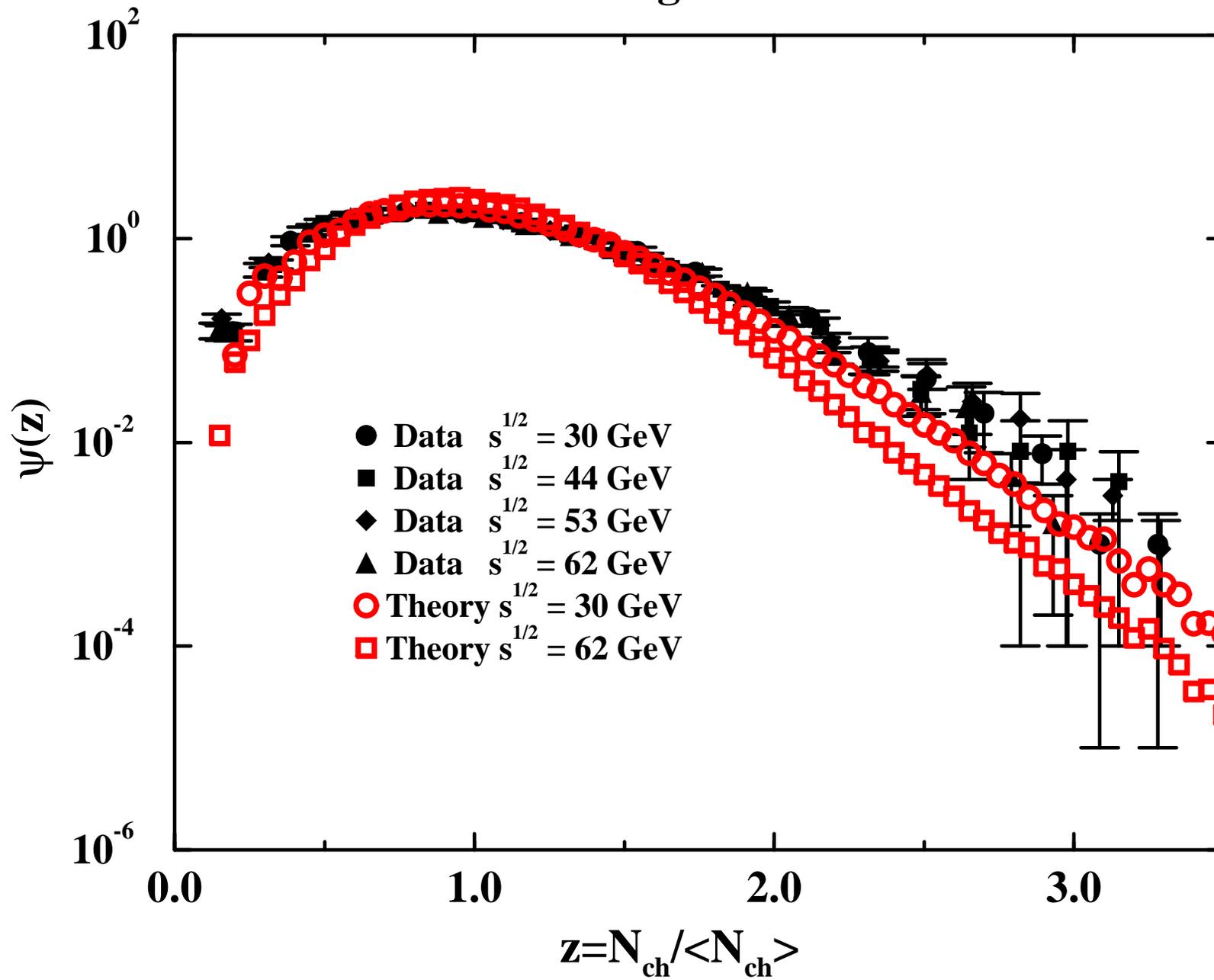
\item[Fig.6: ]  Charged-particle multiplicity distribution in $pp$
collisions at $\sqrt{s}=30$ and $62$ GeV, compared with data \cite{multi}.

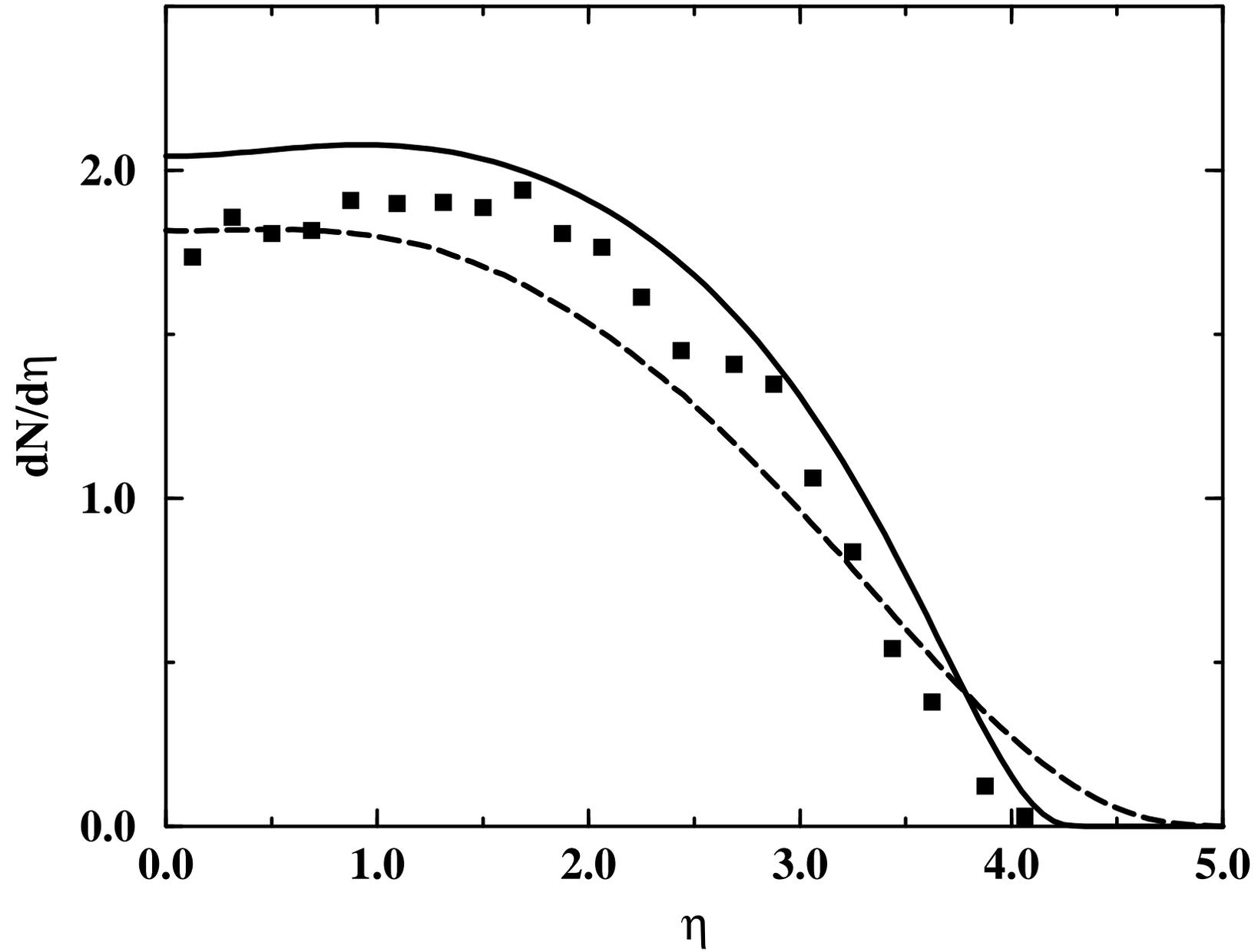
\item[Fig.7: ]  One-particle inclusive pseudo-rapidity distributions for $pp$
collisions at $\sqrt{s}=53$ GeV, computed by using the longitudinal
phase-space model. The solid curve represents the distribution without any
fluctuation, whereas the dashed one is the result with all the fluctuations
included in the way described in the text. The data are from \cite{detaexp}.

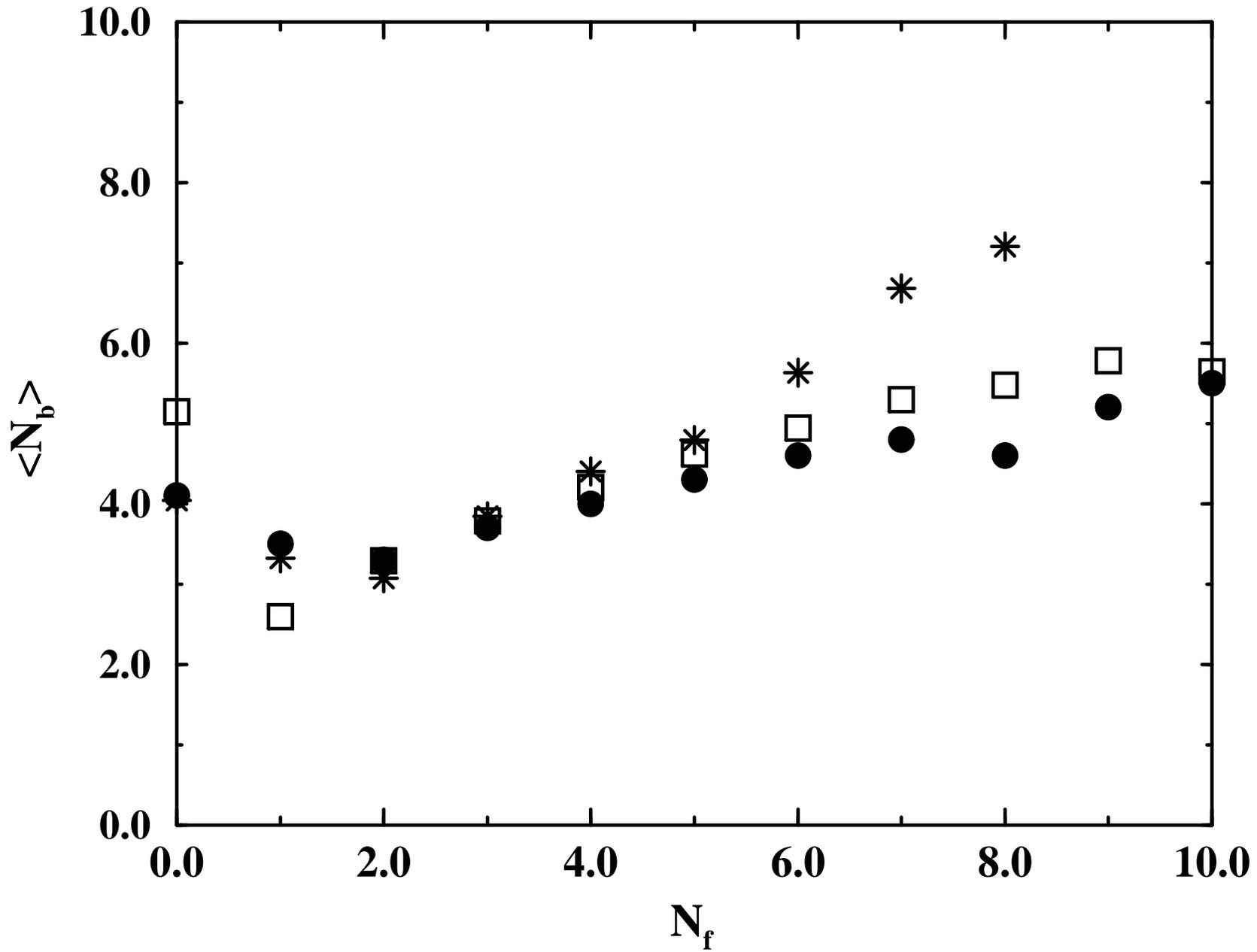
\item[Fig.8: ]  Forward-backward multiplicity correlation in $pp$ collisions
at $\sqrt{s}=24\ $GeV, computed with the longitudinal phase-space model ($%
\Box $), compared with the result without the multiplicity and
final-particle-momentum fluctuations (\hbox{\lower.1cm\hbox{*}}). The data ($%
\bullet $) are from \cite{Uhlig}.
\end{description}

\end{document}